\documentclass[twocolumn]{aastex631}

\begin{document}
\title{\vspace{-0.1in}JWST NIRCam Observations of SN 1987A: Spitzer Comparison and Spectral Decomposition}
\shorttitle{SN 1987A NIRCam Spectral Decomposition}

\author[0000-0001-8403-8548]{Richard G. Arendt} 
\affiliation{Center for Space Sciences and Technology, University of Maryland, Baltimore County, Baltimore, MD 21250, USA}
\affiliation{Code 665, NASA/GSFC, 8800 Greenbelt Road, Greenbelt, MD 20771, USA}
\affiliation{Center for Research and Exploration in Space Science and Technology, NASA/GSFC, Greenbelt, MD 20771, USA}
\email{Richard.G.Arendt@nasa.gov}

\author[0000-0003-4850-9589]{Martha L. Boyer}
\affiliation{Space Telescope Science Institute, 3700 San Martin Drive, Baltimore, MD 21218, USA}

\author[0000-0001-8033-1181]{Eli Dwek} 
\affiliation{Code 665, NASA/GSFC, 8800 Greenbelt Road, Greenbelt, MD 20771, USA}

\author[0000-0002-5529-5593]{Mikako Matsuura}
\affiliation{School of Physics and Astronomy, Cardiff University, Queen's Buildings, The Parade, Cardiff CF24 3AA, UK}

\author[0000-0002-7352-7845]{Aravind P. Ravi} 
\affiliation{Department of Physics, University of Texas at Arlington, Box 19059, Arlington, TX 76019, USA}

\author[0000-0002-4410-5387]{Armin Rest}
\affiliation{Space Telescope Science Institute, 3700 San Martin Drive, Baltimore, MD 21218, USA}
\affiliation{Department of Physics and Astronomy, The Johns Hopkins University, Baltimore, MD 21218, USA}


\author[0000-0002-9117-7244]{Roger Chevalier}
\affiliation{Department of Astronomy, University of Virginia, PO Box 400325, Charlottesville, VA, 22904-4325, USA}

\author[0000-0002-8736-2463]{Phil Cigan}
\affiliation{U.S. Naval Observatory, 3450 Massachusetts Ave NW, Washington, DC 20392-5420, USA}

\author[0000-0001-9419-6355]{Ilse De Looze}
\affiliation{Sterrenkundig Observatorium, Ghent University, Krijgslaan 281-S9, B-9000 Ghent, Belgium}

\author[0000-0001-7906-3829]{Guido De Marchi} 
\affiliation{European Space Research and
Technology Centre, Keplerlaan 1, 2200 AG Noordwijk, Netherlands}

\author[0000-0001-8532-3594]{Claes Fransson}
\affiliation{Department of Astronomy, The Oskar Klein Center, Stockholm University, AlbaNova, 10691 Stockholm, Sweden}

\author[0000-0002-8526-3963]{Christa Gall}
\affiliation{DARK, Niels Bohr Institute, University of Copenhagen, Jagtvej 128, 2200 Copenhagen, Denmark}

\author[0000-0003-1319-4089]{R. D. Gehrz}
\affiliation{Minnesota Institute for Astrophysics, University of Minnesota, 116 Church Street, S. E., Minneapolis, MN 55455, USA}

\author[0000-0003-3398-0052]{Haley L. Gomez}
\affiliation{School of Physics and Astronomy, Cardiff University, Queen's Buildings, The Parade, Cardiff CF24 3AA, UK}

\author[0000-0002-5477-0217]{Tuomas Kangas}
\affiliation{Finnish Centre for Astronomy with ESO (FINCA), FI-20014 University of Turku, Finland}
\affiliation{Tuorla Observatory, Department of Physics and Astronomy, FI-20014 University of Turku, Finland}

\author[0000-0002-3036-0184]{Florian Kirchschlager}
\affiliation{Sterrenkundig Observatorium, Ghent University, Krijgslaan 281-S9, B-9000 Ghent, Belgium}

\author[0000-0002-1966-3942]{Robert P. Kirshner}
\affiliation{TMT International Observatory, 100 West Walnut Street, Pasadena, CA 91124, USA}

\author[0000-0003-0065-2933]{Josefin Larsson}
\affiliation{Department of Physics, KTH Royal Institute of Technology, The Oskar Klein Centre, AlbaNova, SE-106 91 Stockholm, Sweden}

\author[0000-0002-3664-8082]{Peter Lundqvist}
\affiliation{Oskar Klein Centre, Department of Astronomy, Stockholm University, Albanova University Centre, SE-106 91 Stockholm, Sweden}

\author[0000-0002-0763-3885]{Dan Milisavljevic} 
\affiliation{Department of Physics and Astronomy, Purdue University, West Lafayette, IN 47907, USA}

\author[0000-0003-3900-7739]{Sangwook Park} 
\affiliation{Department of Physics, University of Texas at Arlington, Box 19059, Arlington, TX 76019, USA}

\author[0000-0001-5510-2424]{Nathan Smith}
\affil{Steward Observatory, University of Arizona, 933 N. Cherry Avenue, Tucson, AZ 85721, USA}

\author[0000-0001-6815-4055]{Jason Spyromilio}
\affiliation{European Southern Observatory, Karl-Schwarzschild-Str 2, Garching, D-85748, Germany}

\author[0000-0001-7380-3144]{Tea Temim}
\affiliation{Department of Astrophysical Sciences, Princeton University, Princeton, NJ 08544, USA}

\author[0000-0001-7092-9374]{Lifan Wang}
\affiliation{Texas A\&M University, Physics \& Astronomy and  
Mitchell Institute for Fundamental Physics \& Astronomy,
4242 TAMU, College Station, TX 77843-4242, USA}

\author[0000-0003-1349-6538]{J.\ Craig Wheeler}
\affiliation{Department of Astronomy, University of Texas at Austin, Austin, TX}

\author[0000-0001-6567-627X]{Charles E. Woodward}
\affiliation{Minnesota Institute for Astrophysics, University of Minnesota, 116 Church Street, S. E., Minneapolis, MN 55455, USA}

\begin{abstract}
JWST NIRCam observations at $1.5-4.5$ $\mu$m have provided broad and narrow band imaging of the 
evolving remnant of SN 1987A with unparalleled sensitivity and spatial resolution.
Comparing with previous marginally spatially resolved Spitzer IRAC observations from $2004-2019$ 
confirms that the emission arises from the circumstellar equatorial ring (ER),
and the current brightness at 3.6 and 4.5 $\mu$m was accurately predicted by extrapolation 
of the declining brightness tracked by IRAC. 
Despite the regular light curve, the NIRCam observations 
clearly reveal that much of this emission is from a newly developing
outer portion of the ER. Spots in the outer ER tend to lie at position angles in between  
the well-known ER hotspots.
We show that the bulk of the emission in the field can be represented by 5 standard 
spectral energy distributions (SEDs), each 
with a distinct origin and spatial distribution. 
This spectral decomposition provides a powerful technique for distinguishing 
overlapping emission from the circumstellar medium (CSM) and the supernova (SN) 
ejecta, excited by the forward and reverse shocks respectively. 
\end{abstract}


\section{Introduction} \label{sec:intro}
The last thirty six years of observations of SN 1987A have revealed a wealth of information on the evolution, composition, dynamics, and morphology of the supernova (SN) ejecta and its surrounding circumstellar medium (CSM). 
SN 1987A was observed at almost all wavelengths from gamma ray to radio, revealing 
the underlying physical processes of the SN explosion and its aftermath, as 
reviewed by \cite{Arnett:1989}, \cite{McCray:1993, McCray:2007}, and 
\cite{McCray:2016}.
Most relevant to the current analysis are the advances in our understanding of the equatorial ring (ER) 
around the SN, the origin of its infrared (IR) emission, and the interaction of the ER with the advancing SN blast wave.

SN 1987A is the first supernova for which the CSM from the progenitor star can be spatially resolved and the transition from SN to supernova remnant (SNR) can be observed
with modern instrumentation.
The densest concentration of the CSM 
was found to be an ER, rather than a spherical shell.
This was first inferred from the presence of narrow UV and optical emission lines from gas that was flash ionized and excited by the UV flash generated by the shock breakout through the surface of the progenitor star  
\citep{Fransson:1989,Fransson:1989a,Wood:1989,Lundqvist:1991,Dwek:1992b}. 
As the glare of the SN faded, the first images of this ring structure were obtained about three years 
after the explosion \citep{Wampler:1990, Panagia:1991} and two fainter ``outer'' rings centered north and south 
of the SN were also revealed.

Measurements of the photo-ionized ER after the SN explosion indicated 
a projected semimajor axis of $0\farcs82$ ($ = 0.20$ pc $= 0.65$ ly at a distance 
of 50 kpc), and an inclination of $\sim43\arcdeg$ with the north side being nearer
\citep{Jakobsen:1991,Panagia:1991,Plait:1995,Burrows:1995}. There is evidence that the ER is not quite
intrinsically circular, with $b/a = 0.98$ \citep{Sugerman:2005}.

Very similar equatorial rings have been seen around some other 
blue supergiants, with similar structure in the equatorial clumps and 
mild eccentricity 
\citep[e.g.,][]{Brandner:1997, Smith:2007, Smith:2013}.
Rings like this might arise at the interaction between a BSG wind and a preceding RSG wind \citep{Blondin:1993,Martin:1995,Chevalier:1995},
or from the ejecta in a stellar merger event \citep{Morris:2009}.

The direct interaction of the SN shock with the ER manifested itself in 1995 with the first appearance of 
bright optical hotspots (or knots) on the north-east quadrant of the ER 
\citep{Sonneborn:1998,Lawrence:2000}. 
The hotspots are clumps of dense gas that are lit up as the shock propagates into them.  With the 
progression of the shock more hotspots appeared to fully encircle the ring by 2005 
\citep{Fransson:2015, Kangas:2023}. About 4 years later 
(day $\sim8000$), the total brightness of the ER was fading at 
most visible wavelengths \citep{Fransson:2015}.

The ER was first imaged at mid-IR wavelength with the Thermal-Region Camera 
Spectrograph \citep[T-ReCS;][]{Telesco:1998} on the 8.1 m Gemini-South Telescope by \cite{Bouchet:2004}. 
Photometric light curves at 10 and 20 $\mu$m showed that around day 4000 the dominant contribution to the SN energy output transitioned from the ejecta to the shock-ER interaction. 
The mid-IR emission ($3-40$ $\mu$m) arises from pre-existing ER
dust that is collisionally heated by the shocked X-ray emitting gas. 

The Spitzer \citep{Werner:2004, Gehrz:2007} Infrared Spectrograph \citep[IRS,][]{Houck:2004} 
$5-30$ $\mu$m spectrum of the ER showed that the emitting dust consists of $\sim$ 160-180~K silicate grains
with radii of $a\gtrsim 0.2$ $\mu$m \citep{Bouchet:2006, Dwek:2010}.
The IRS spectrum also showed an excess of 5-8 $\mu$m emission over that of the silicate dust, a spectral component that has been attributed to emission from very small, $a < 0.03$ $\mu$m, and hot, $T \sim350-500$~K, grains with several candidate compositions (Dwek et al. 2010). Assuming that the two dust components reside in the same shocked gas, the small hot grains should have lifetimes that are about 10 times shorter than the larger silicate grains.
\cite{Jones:2023} use new James Webb Space Telescope \citep[JWST;][]{Gardner:2023}
observations to address the 
issue of small hot grains. Notably they find that the full mid-IR spectrum 
can be modeled with a single grain composition, but multiple temperatures
are required.

Photometric mid-IR light curves obtained with Spitzer showed a similar evolutionary behavior, manifested by an approximately constant 24 to 3.6 $\mu$m flux ratio up to day $\sim$~8000, which marked the end of the Spitzer cryogenic era. 
Subsequent warm-era Spitzer photometric observations showed that the 3.6 and 4.5 $\mu$m light curves peaked around day 8000 and started to decline after day $\sim$ 8500 \citep{Arendt:2016, Arendt:2020}. 
This behavior is also exhibited by ground-based $J$, $H$, and $K_s$ measurements \citep{Kangas:2023}.
\cite{Arendt:2020} showed that these light curves can be modeled as the product of a convolution of a Gaussian and an exponential function. 

The 3.6 and 4.5 $\mu$m light curves are generated by the small hot grains. 
Due to grain sputtering, this emission is expected to drop significantly below the 
mid-IR light curve when the age of the shocked gas exceeds the lifetime of the small grains, 
but is still shorter than the lifetime of larger silicate grains
\citep{Dwek:1996, Dwek:2008}. 
The lack of any mid-IR photometry beyond day 8000 
prevented this prediction from being tested with Spitzer, but 
the JWST Mid Infrared Instrument (MIRI) Medium Resolution Spectrometer (MRS) \citep[][]{Wright:2023}
observations indicate that the relative brightness of the 
hot dust has decreased with respect to the cooler dust \citep{Jones:2023}.

The high-resolution JWST 
Near Infrared Camera \citep[NIRCam;][]{Rieke:2023}
broad and narrow band images at 1.5 to 4.5 $\mu$m presented in this paper allow us to address several issues raised by previous observations, namely the consistency between Spitzer IRAC and the JWST NIRCam images and photometry, the validity of the light curves model at later epochs, the origin of the hot emission component, the interaction of the blast wave with the ER, and variations in the spectral energy distribution (SED) of the various spatial components of the ER and the immediate surrounding medium.

We review the JWST NIRCam data and its processing in Section~2. 
In Section~3 we compare the NIRCam data with the Spitzer 
Infrared Array Camera \citep[IRAC;][]{Fazio:2004} imaging and photometry.
The much lower resolution IRAC images had only hinted at the complex structure of the region. Deconvolution of these images \citep{Arendt:2020} to a 0.2\arcsec\ resolution suggested that the region is comprised 
of three stars and the ER, with no 
noticeable emission from the ejecta at 3.6 and 4.5 $\mu$m. The newly obtained high resolution JWST NIRCam images validate the results of the deconvolution and show that there were no major morphological changes between the Spitzer and JWST eras. 
The JWST observations enable the use of aperture photometry to obtain a more accurate determination of the brightness of the ER, 
separate from the confusion of the nearby stars. 
The NIRCam F356W and F444W photometry is used to extend the 16-year long 
light curves measured by Spitzer by another 3 years.

In Section 4 we take a close look at extended emission (brightest in the 
F444W band) beyond the well-known hotspots of the ER. Structure in this new 
outer ER region is found to be anti-correlated with the ER hotspots, an apparent 
imprint on the structure of shocks that have swept around and past the hotspots. 

In Section 5 we use selected SED templates to decompose the 
NIRCam images into different spatial structures associated with each of 
the templates. This technique can separate the structures of 
overlapping components, and residuals provide indication of where 
additional or modified components may exist.

Our results are summarized in Section 6.
      
\section{Data} \label{sec:data}
The observations\footnote{JWST Proposal ID 1726 -- PI: M. Matsuura}
(on day 12975 since the explosion of SN 1987A) 
comprise NIRCam images in wide bands: F150W, F200W, F356W, F444W
and narrow bands: F164N, F212N, F323N, F405N.
All the JWST NIRCam data used in this paper can be found in MAST: 
\dataset[10.17909/dzkq-7c90]{http://dx.doi.org/10.17909/dzkq-7c90}.
The wide bands provide the overall SED of the spatially resolved emission. 
The selected narrow band filters are designed to isolate specific spectral 
lines of H$_2$ (F212N, F323N), \ion{H}{1} (F405N), 
and a blend of \ion{Fe}{2} and \ion{Si}{1} (F164N).
With JWST, we now have the ability to spatially resolve the locations of these species 
in great detail.

The images were processed 
to use improved calibration,
astrometry, and subtraction of artifacts and correlated noise with 
respect to the standard pipeline results (in the MAST archive\footnote{\url{https://archive.stsci.edu/}}).
We used the WCS alignment tool JHAT \citep{Rest:2023} to align the NIRCam images to Gaia DR2.
To remove residual striping, in this work median values per row are
subtracted for each detector. To remove the overall background, 
the median value of each image is subtracted as well.
When used for the spectral decomposition (in Section 5), 
all of the shorter wavelength images are additionally convolved 
to the $0\farcs145$ full width half maximum (FWHM) resolution of the F444W image.
See \cite{Matsuura:2023} for further details.

\begin{figure*}[ht] 
   \centering
   \includegraphics[width=7in]{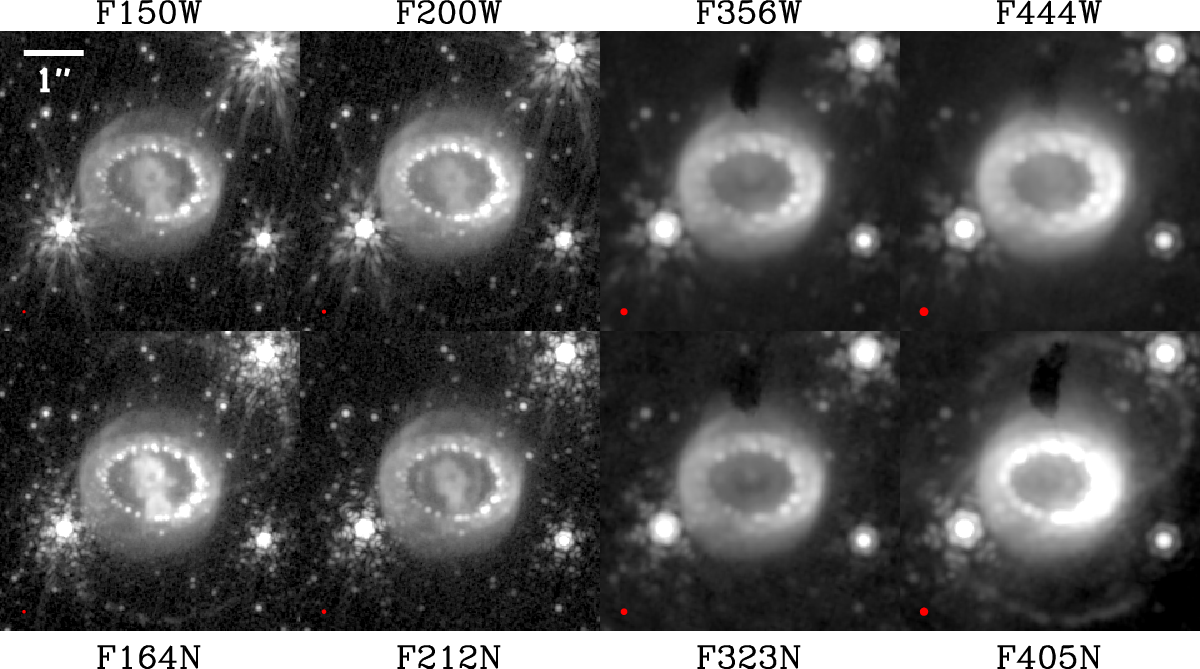}\\ 
   ~~\\
   \includegraphics[width=7in]{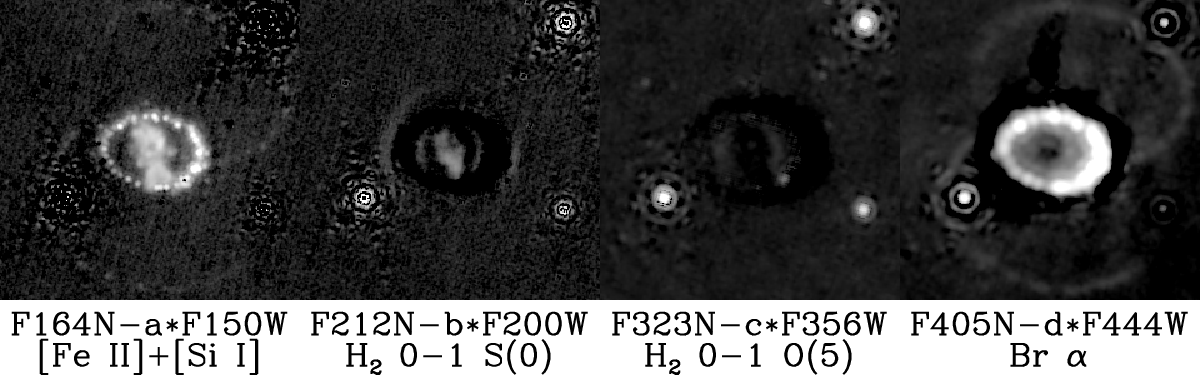} 
   \caption{NIRCam images of SN 1987A reprojected onto a common ($\alpha$, $\delta$) coordinate grid. 
   The top row contains the broad band images. The middle row shows narrow band 
   images that are respectively centered on lines for: [\ion{Fe}{2}]+[\ion{Si}{1}], H$_2$, H$_2$ (also), and \ion{H}{1}. 
   The images are logarithmically scaled from 0.32 to 32 MJy sr$^{-1}$ (after adding an offset of 0.5 MJy sr$^{-1}$),
   and show a field of view of $5''\times5''$. 
   North is up and east is to the left in all images in this paper.
   Red dots in the lower left corner of each panel indicate 
   the FWHM size of the point spread function (PSF).
   Relative brightness variations between the inner ejecta, the hotspots 
   of the ER, and the emission beyond the ER hotspots indicate a variety of emission mechanisms at work.
   The bottom row emphasizes the line emission in each of the narrow bands by 
   subtracting scaled versions of the wide band emission.}
   \label{fig:images}
\end{figure*}

\begin{figure}
\begin{interactive}{animation}{SN1987A_wave.mp4}
\includegraphics[width=3.0in]{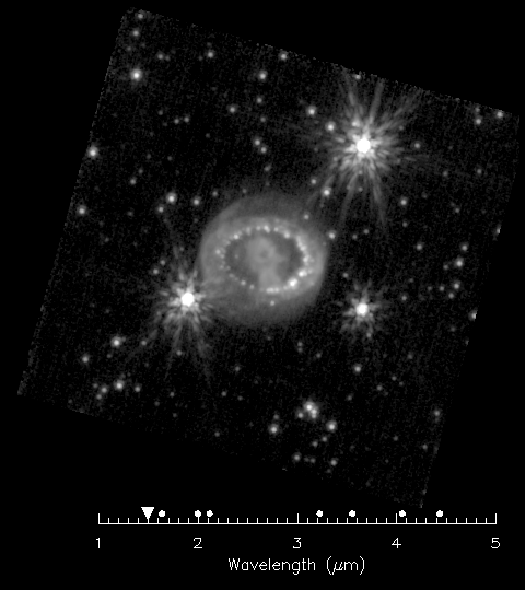}
\end{interactive}
\caption{Animated comparison of the SN 1987A emission in 
the different NIRCam bands. Showing the data of Figure \ref{fig:images} 
as a single image
while smoothly transitioning between wavelengths 
reveals similarities and differences between the structures at 
different wavelengths. As the 18-second animation plays, the 
scale bar and pointer at the bottom indicate which wavelength, or intermediate
blend of wavelengths, is displayed.}
\label{fig:animation_gray}
\end{figure}

\begin{figure}
\begin{interactive}{animation}{SN1987A_wave_color.mp4}
\includegraphics[width=3.0in]{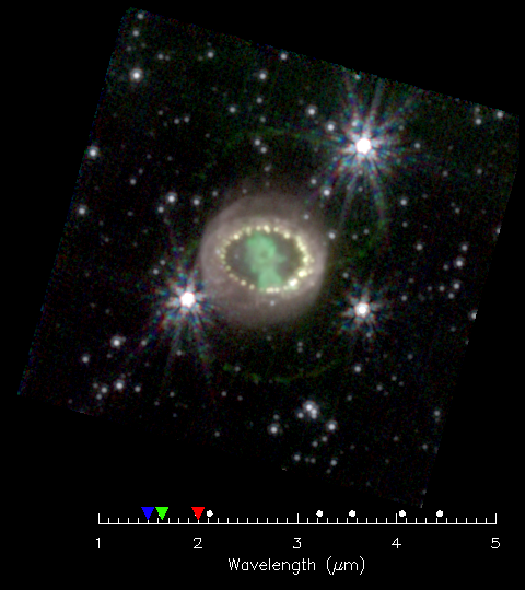}
\end{interactive}
\caption{Animated 3-color comparison of the SN 1987A emission in 
the different NIRCam bands. Similar to Figure \ref{fig:animation_gray}, 
this animation helps highlight 
color differences between different structures. 
As the 13-second animation plays, the 
scale bar and pointers at the bottom indicate which 
wavelength, or intermediate blend of wavelengths, is displayed
in each of the red, green, and blue channels.}
\label{fig:animation_color}
\end{figure}

The images are reprojected from their intrinsic orientation and
scale to a common grid in $(\alpha,\delta)$ coordinates, with the pixel
scale of the short wavelength images. These are shown in 
Figure \ref{fig:images}. 
Figures \ref{fig:animation_gray} and \ref{fig:animation_color} 
are grayscale and 3-color animations of these
eight images that provide an alternate way of comparing the brightness, 
structure, and color of the emission at different wavelengths.

We also show emission line images in Figure \ref{fig:images}
which are constructed by subtracting scaled versions of the corresponding 
wide band images. In principle, this can remove the continuum contribution 
in the narrow band images. However, the process is imperfect because the 
continuum in the wide band images has variations in color, and because 
the wide band images contain (diluted) contributions from the targeted 
emission lines as well as other lines within the wider bandwidths.

\section{Comparison With Spitzer IRAC}

SN 1987A had been monitored at roughly 6 month intervals 
throughout the Spitzer mission in order to trace the evolution 
of the SN \citep{Bouchet:2006, Dwek:2008, Dwek:2010, Arendt:2016, Arendt:2020}. 
During Spitzer's cryogenic mission the observations used all 
instruments, but after the helium ran out in May 2009, 
observations through Sep 2019 used only IRAC imaging at 3.6 and 4.5 $\mu$m.
These bands are very similar to the F356W and F444W bands of NIRCam.

\begin{figure}[tbp] 
   \centering
   \includegraphics[width=3.25in]{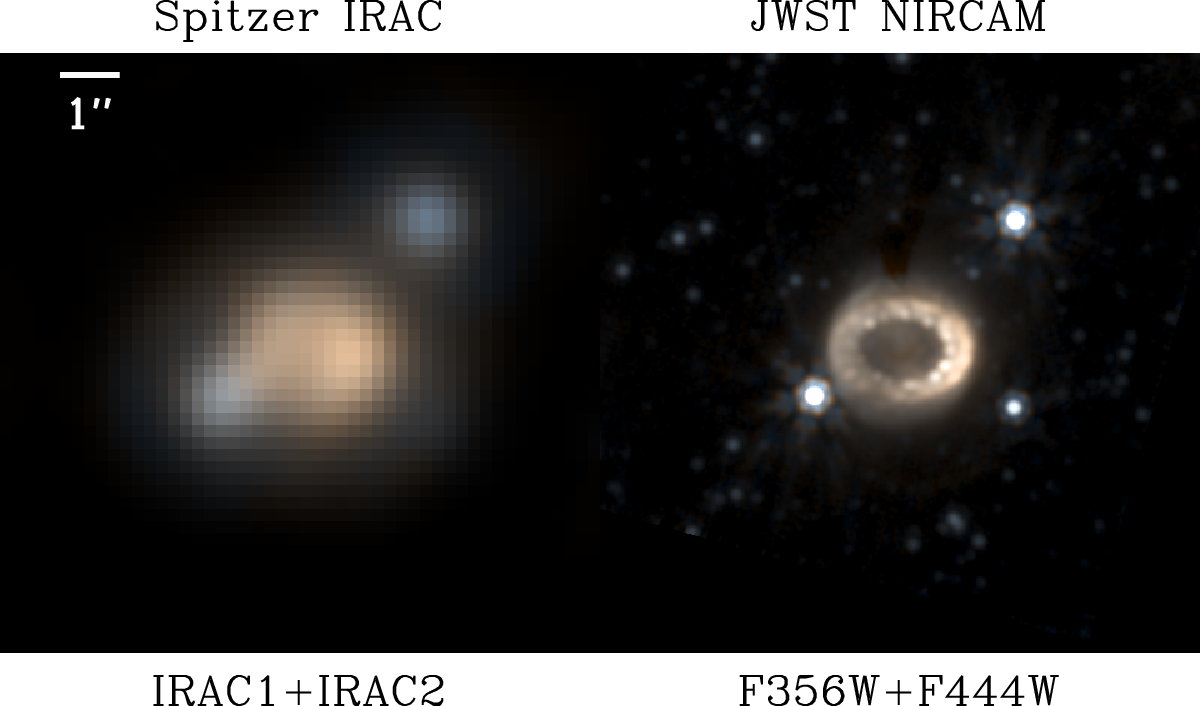} 
   \caption{Comparison of IRAC and NIRCam images at comparable wavelengths 
   provides a qualitative indication that the high resolution mapping and 
   deconvolution of the IRAC data \citep{Arendt:2020}
   had revealed hints of the true structure of the SNR emission, and that there 
   have been no major changes between the Spitzer and JWST eras.}
   \label{fig:IRACcolor}
\end{figure}

\begin{figure*}[h] 
   \centering
   \includegraphics[height=2.25in]{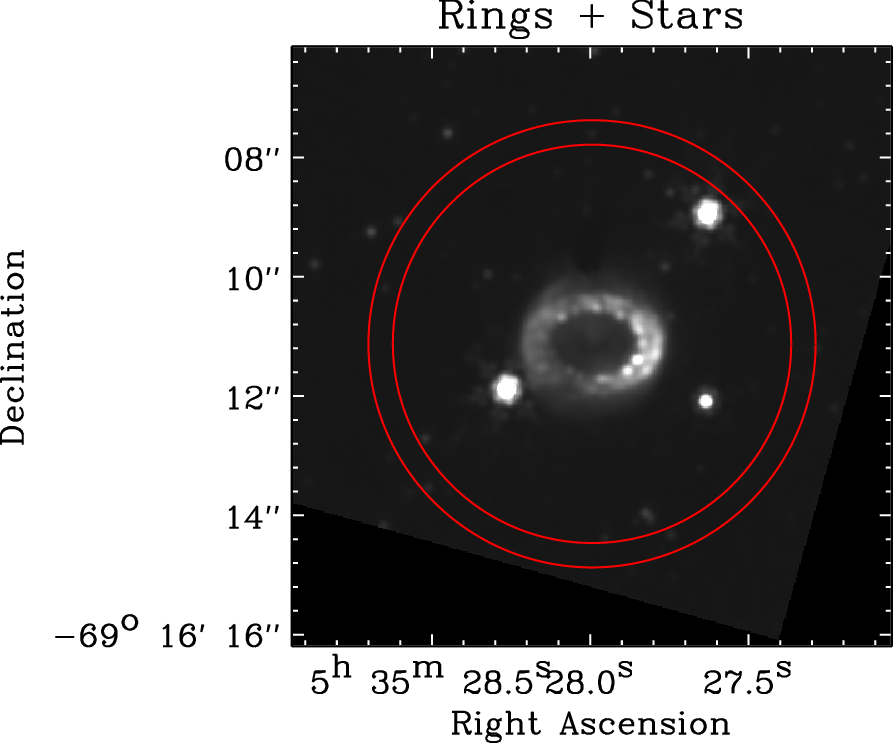}~~ 
   \includegraphics[height=2.25in]{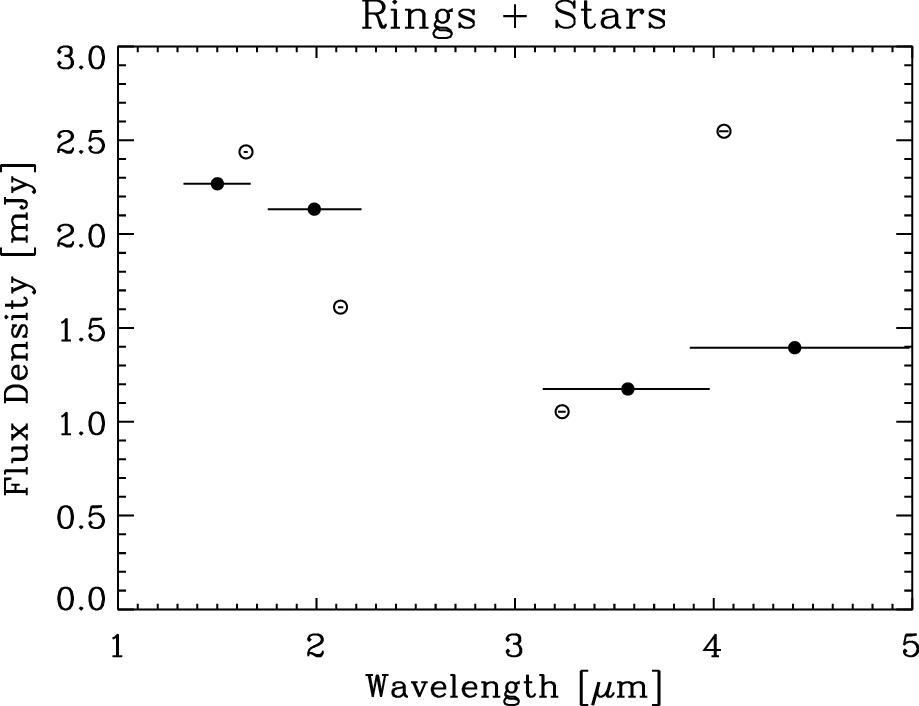}\\ 
   ~~\\
   \includegraphics[height=2.25in]{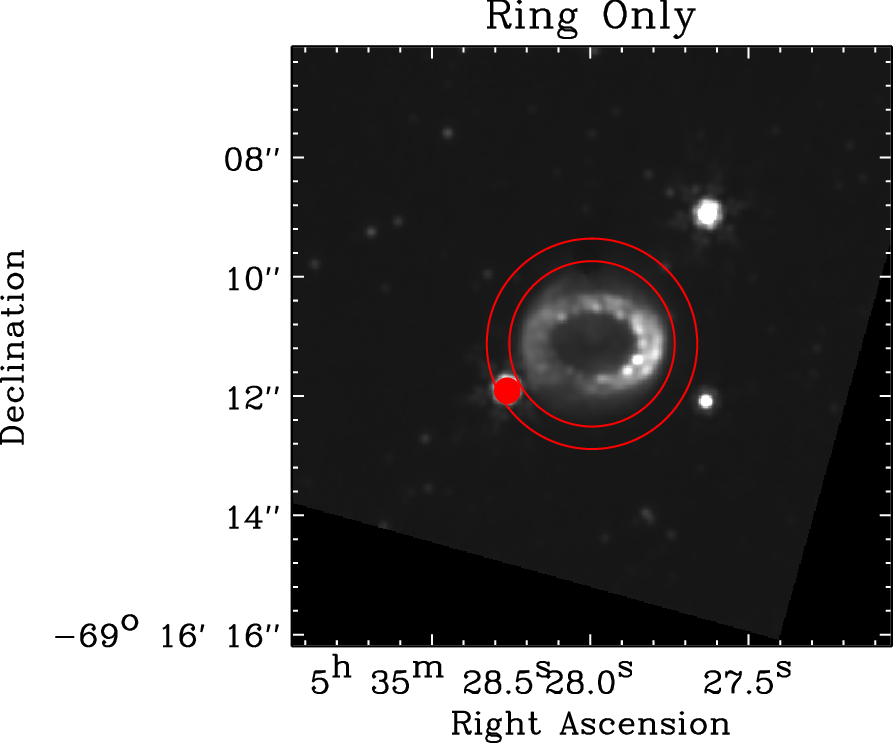}~~
   \includegraphics[height=2.25in]{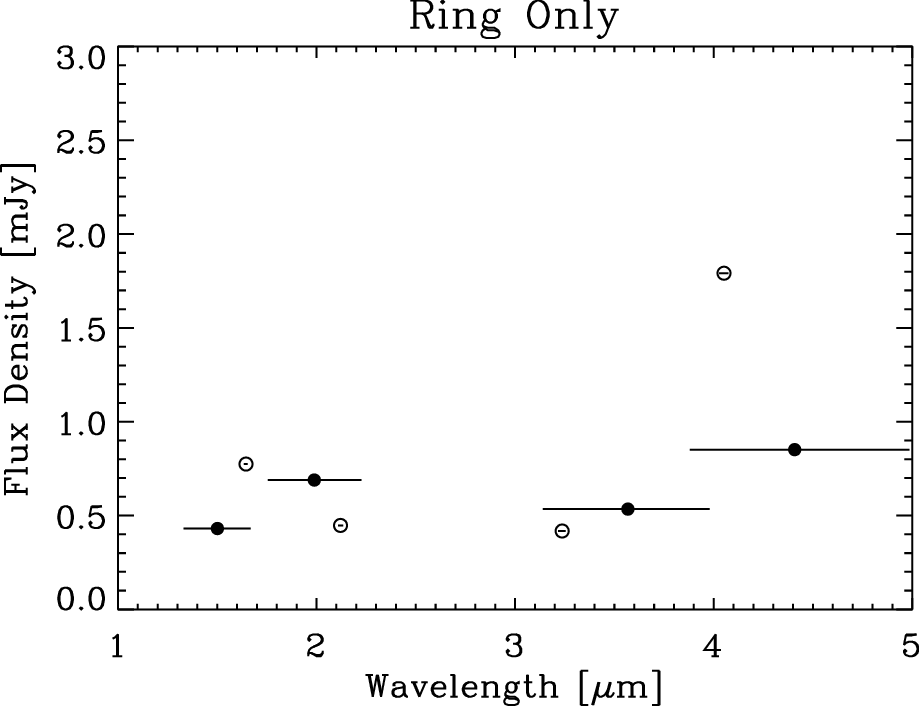}\\ 
   ~~\\
   \includegraphics[height=2.25in]{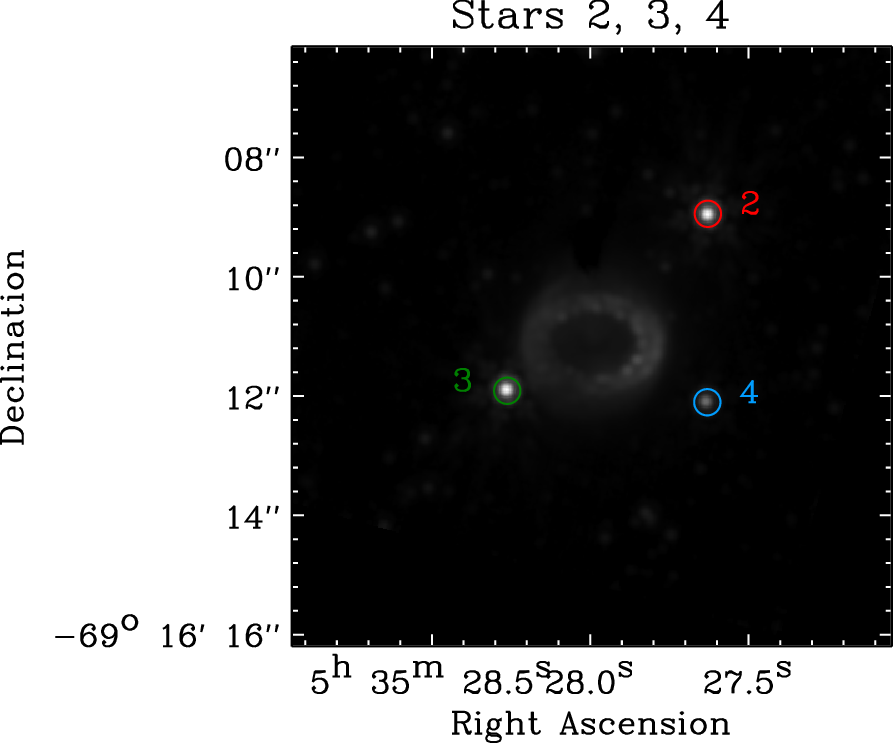} ~~
   \includegraphics[height=2.25in]{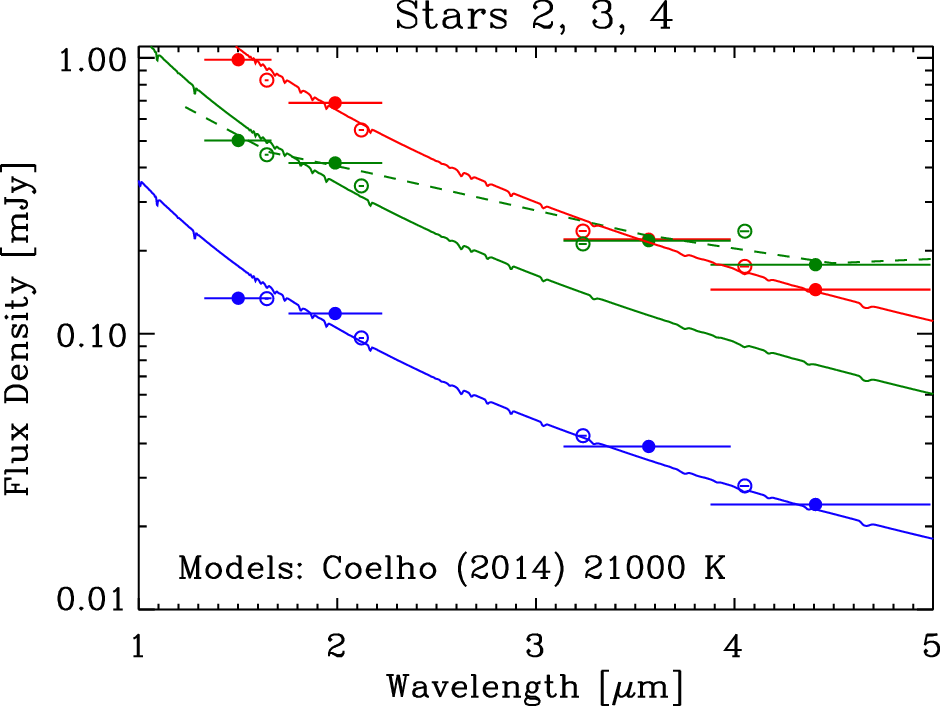} 
   \caption{The top row shows (left) apertures used to replicate IRAC
   photometry that had been performed on much lower resolution data,
   and (right) the resulting spectral energy distribution (SED). These
   flux densities are the sum of those for the SN and stars 2, 3, and 4.
   Horizontal lines indicate the bandwidths for wide bands (filled circles)
   and narrow bands (open circles).
   The middle row shows results with a smaller aperture that includes 
   only the SN. Star 3 is specifically excluded from the background region.
   The bottom row shows the SEDs of stars 2, 3, and 4, with 21000K 
   stellar atmosphere models from \cite{Coelho:2014} shown for comparison.
   Star 3 is a classical Be star with excess emission at $>3$ $\mu$m. 
   The green dashed line shows the SED of the B2IIIe star 
   SMC5\_074402 from \cite{Bonanos:2010} for comparison (multiplied by a factor of 0.8).
   The display range on the bottom left panel is altered such that the stars are not
   saturated.} 
   \label{fig:photometry}
\end{figure*}

Figure \ref{fig:IRACcolor} shows a comparison of the 3.6 and 4.5 $\mu$m
imaging available from IRAC versus that from NIRCam. The IRAC 
imaging shown here \citep{Arendt:2020} used deconvolution techniques to achieve sufficient
resolution to barely resolve the ER. The comparison validates that deconvolution, 
which showed that the west side of the ER was brighter than the east, 
and separated out stars 2, 3, and 4. [Note that the designation of star 4
\citep[``Star A'' in][]{Kangas:2023}, 
the brightest one to the southwest (Figure \ref{fig:photometry}), follows \cite{Arendt:2020} and 
this is not the same as star 4 of \cite{Walker:1990}.]

With normal processing of IRAC data at single epochs, the SN was not resolved 
from stars 2, 3, and 4. Thus aperture photometry measured the combined 
fluxes of the SN and these stars, and estimates of the stellar fluxes based 
on extrapolation of $JHK$ photometry were subtracted to obtain the SN brightness.
For comparison with the IRAC measurements we performed aperture 
photometry on the NIRCam images using large apertures that replicated 
those used for IRAC. These apertures and the resulting flux densities are
shown in the top row of Figure \ref{fig:photometry}. The second row
shows similar integrated photometry from the SN alone with a smaller aperture 
that excludes stars 2, 3, and 4. Special masking (indicated in the Figure)
was applied so that star 3 did not affect the background measurements. 
The last row
of Figure \ref{fig:photometry} shows aperture photometry for stars 2, 3, and 4, 
individually. Star 2 is well represented by a 21000 K model atmosphere, 
as is appropriate for an early B type star. Star 4 appears to be slightly 
cooler, as might be anticipated by its lower brightness. Star 3 is 
a known classical Be star \citep{Wang:1992, Walborn:1993} and shows
excess emission at $\lambda > 3$ $\mu$m which is typical of these stars.
The photometry is listed in Table 1.

\begin{deluxetable*}{lccccc}
\tablecaption{Flux Densities}
\tablewidth{0pt}
\tablehead{
\colhead{} & 
\colhead{} & 
\colhead{} & 
\colhead{} & 
\colhead{SN 1987A} & 
\colhead{SN 1987A}\\ 
\colhead{Filter} & 
\colhead{Star 2} & 
\colhead{Star 3} & 
\colhead{Star 4} & 
\colhead{Small Ap.} &
\colhead{Large Ap.}
}
\startdata
F150W &  $0.985\pm0.001$ &  $0.502\pm0.001$ &  $0.134\pm0.001$ &  $0.431\pm0.001$ &  $2.268\pm0.001$\\
F164N &  $0.830\pm0.001$ &  $0.445\pm0.001$ &  $0.134\pm0.001$ &  $0.775\pm0.001$ &  $2.438\pm0.001$\\
F200W &  $0.687\pm0.001$ &  $0.416\pm0.001$ &  $0.118\pm0.001$ &  $0.689\pm0.001$ &  $2.133\pm0.001$\\
F212N &  $0.548\pm0.001$ &  $0.343\pm0.001$ &  $0.097\pm0.001$ &  $0.447\pm0.001$ &  $1.611\pm0.001$\\
F323N &  $0.235\pm0.001$ &  $0.212\pm0.001$ &  $0.043\pm0.001$ &  $0.418\pm0.001$ &  $1.054\pm0.001$\\
F356W &  $0.220\pm0.001$ &  $0.217\pm0.001$ &  $0.039\pm0.001$ &  $0.535\pm0.001$ &  $1.175\pm0.001$\\
F405N &  $0.175\pm0.001$ &  $0.235\pm0.001$ &  $0.028\pm0.001$ &  $1.792\pm0.003$ &  $2.548\pm0.002$\\
F444W &  $0.144\pm0.001$ &  $0.178\pm0.001$ &  $0.024\pm0.001$ &  $0.851\pm0.002$ &  $1.395\pm0.001$\\
\enddata
\tablecomments{units = mJy}
\end{deluxetable*}

\begin{figure}[tbp] 
   \centering
   \includegraphics[height=2.25in]{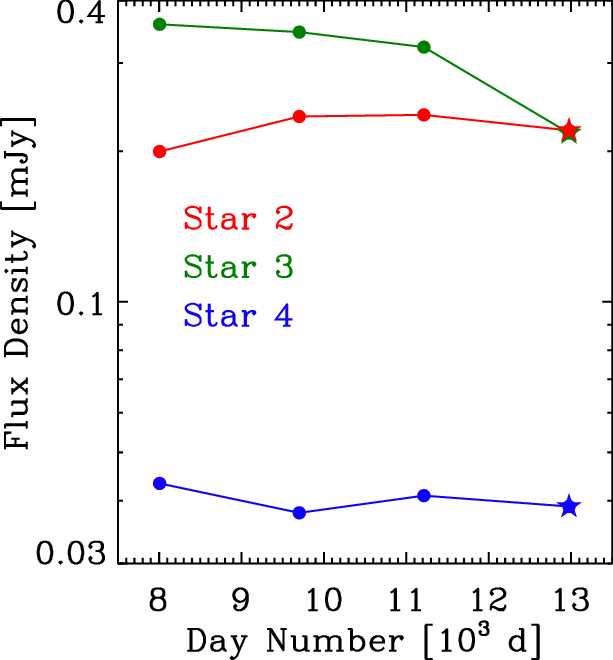} 
   \caption{3.6 $\mu$m stellar photometry (light curves) of stars 2, 3 and 4. 
   The filled circles are modeled brightnesses from IRAC data 
   \citep{Arendt:2020}, where the stars are not fully resolved. 
   The filled star symbols at day 12975 indicate NIRCam measurements. 
   Stars 2 and 4 appear to be stable, while star 3 (a classical Be star) 
   appears to show a decreased brightness. Star 3 is intrinsically variable, 
   but, in addition, the IRAC flux densities may have been affected by confusion with the ER.}
   \label{fig:stars}
\end{figure}

Figure \ref{fig:stars} compares the modeled (not directly measured) photometry
of these stars with IRAC \citep{Arendt:2020} at 3.6 $\mu$m with the NIRCam 
F356W measurement. The results for stars 2 and 4 indicate that 
the IRAC and NIRCam results (at 3.6 $\mu$m) are consistent and 
that these stars are stable to within $\sim10\%$. Star 3 shows a
drop in brightness from IRAC to NIRCam measurements. This 
may be intrinsic variability of the Be star, but it is also possible that, 
being the closest star to the ER, its IRAC flux density estimates were 
contaminated by emission from the ER. Future NIRCam observations
of SN 1987A will definitively reveal the intrinsic variability of Star 3
at the NIRCam wavelengths \citep[cf.][]{Walborn:1993}. 

\begin{figure}[tbp] 
   \centering
   \includegraphics[width=3.5in]{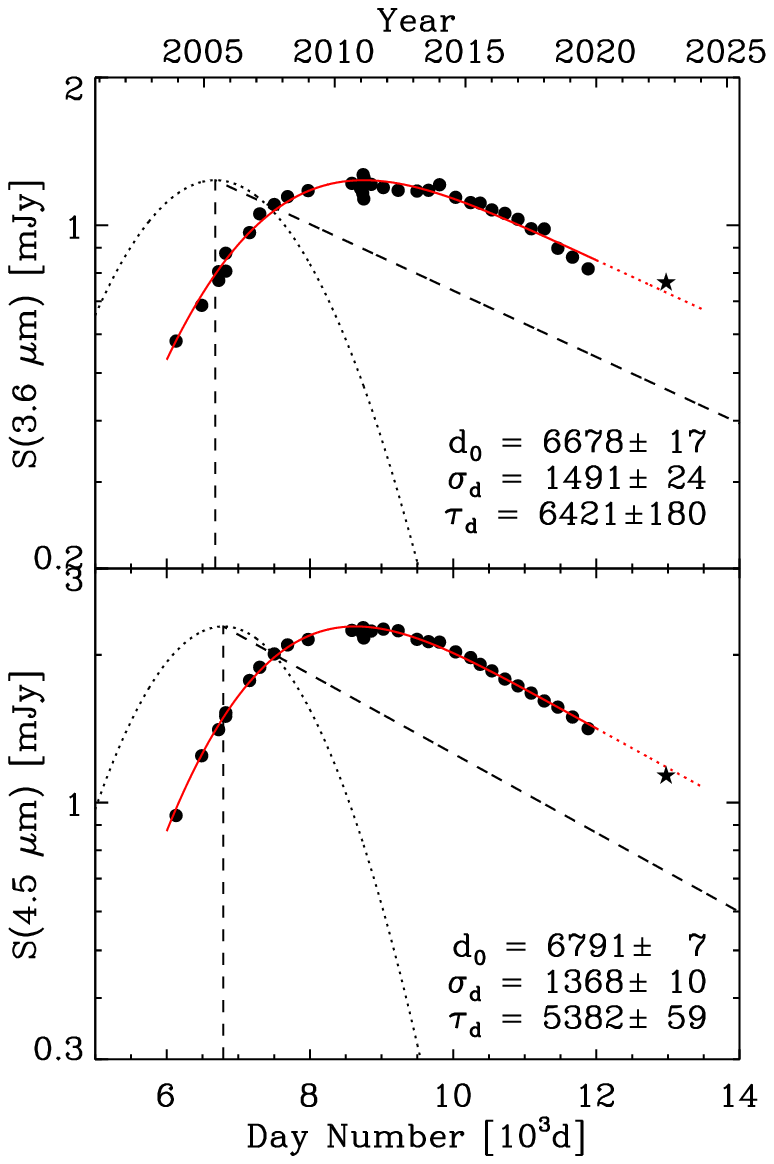} \\
   \caption{The IRAC light curves at 3.6 and 4.5 $\mu$m are shown with 
   filled circles. The NIRCam measurements are indicated with filled stars 
   at day 12975. Formal random uncertainties are smaller than the plotted symbols.
   For consistency with the IRAC 
   measurements these are the large aperture flux densities 
   minus 0.41 and 0.26 mJy estimates for the 3.6 and 4.5 $\mu$m 
   combined brightness of Stars 2 and 3. The dashed red line shows that an extrapolation of 
   the IRAC model \citep[solid red line,][]{Arendt:2020} is a good predictor of the NIRCam flux 
   densities. The IRAC models are the convolution of a Gaussian function 
   and an exponential function (shown as black curves at $d_0$).}
   \label{fig:model19}
\end{figure}

During the Spitzer cryogenic mission, the brightness of SN 1987A
had been steadily increasing. Continued monitoring during
the warm mission showed the brightness peaking and declining 
at 3.6 and 4.5 $\mu$m. \cite{Arendt:2020} found that these 
light curves could be well modeled as the convolution of a 
Gaussian function and an exponential decay term. One 
interpretation of this is that the Gaussian function represents 
the column density distribution of material swept up in the ER
(combined with light travel times from the near to far side of the ER),
and the exponential decay represents the temporal evolution of the emissivity of
each small parcel of gas and dust, starting from the moment it is shocked. 
In Figure \ref{fig:model19} we show the IRAC light curves and 
the original models. We have added the new NIRCam measurements
and extrapolated the models to the time of the new observations. 
At both wavelengths the extrapolations are within a few percent 
of the NIRCam measurement, indicating that the empirical model 
continues to serve as a good predictor of the SNR brightness.
However, there is some hint of flattening of the 3.6 $\mu$m light curve if 
one compares the final IRAC epochs with the NIRCam measurement.
This may reflect unrecognized systematic errors in the 
late Spitzer 3.6 $\mu$m measurements, which are more susceptible to noise and 
to contamination from Stars 2 and 3 than 4.5 $\mu$m measurements.
However, if it is a real trend borne out future JWST observations,
this may reflect the development of the reverse shock (RS) structure 
beyond the knotty ER. The brightest location of the
reverse shock to the northeast of the ER exhibits a bluer SED  
than that of the outer ER (seen in Figures \ref{fig:samples} and 
\ref{fig:compare321} in Section 5.1).

\clearpage
\section{Structure of the Equatorial Ring}

\begin{figure}[tbp] 
   \centering
   \includegraphics[width=1.65in]{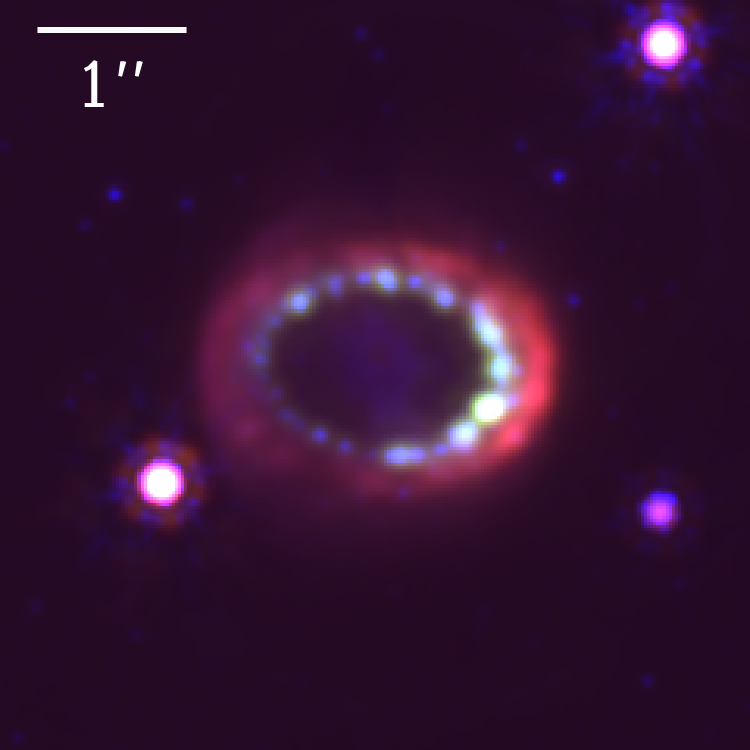}
   \includegraphics[width=1.65in]{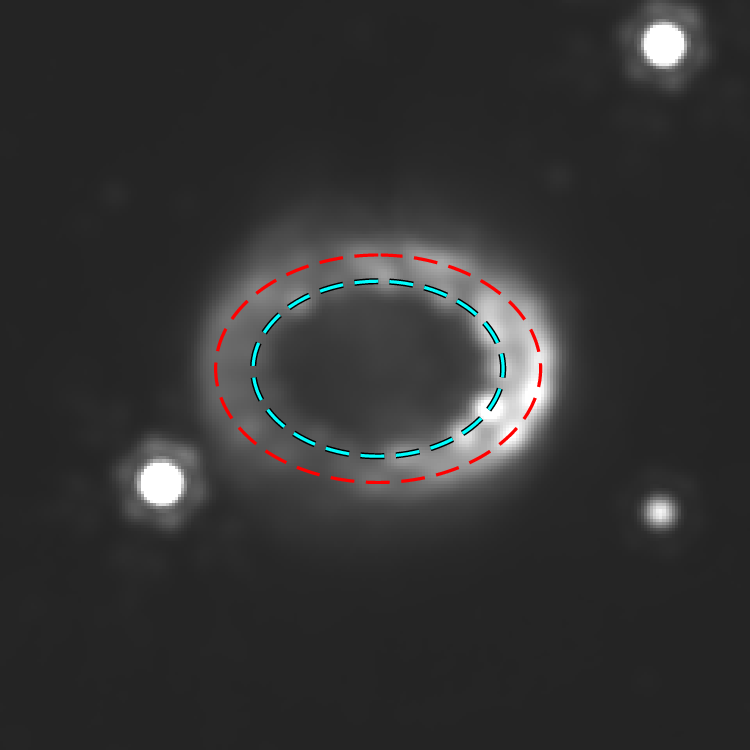}
   \caption{Locations of ER hotspots and clumps. Left: Images at F444W (red), F405N (green), and F200W (blue). 
   Right: F444W image superimposed with ellipses tracing the bright hotspots ( blue) and 
   the outer emission (red) as shown in Figure \ref{fig:er}.}
   \label{fig:er_rings}
\end{figure}

\begin{figure}[ht] 
   \centering
   \includegraphics[height=1.90in]{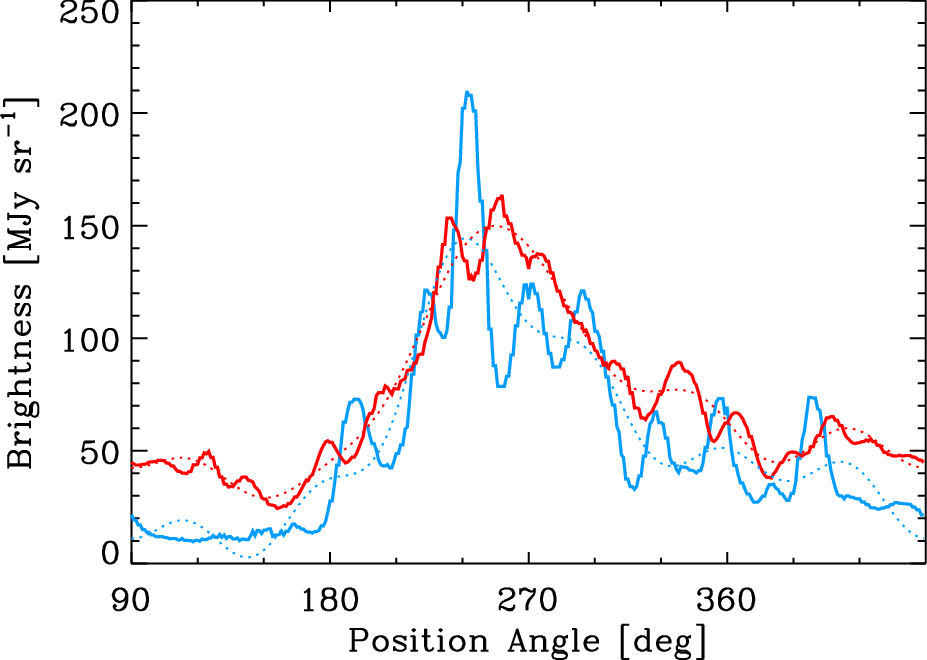}
   \includegraphics[height=1.90in]{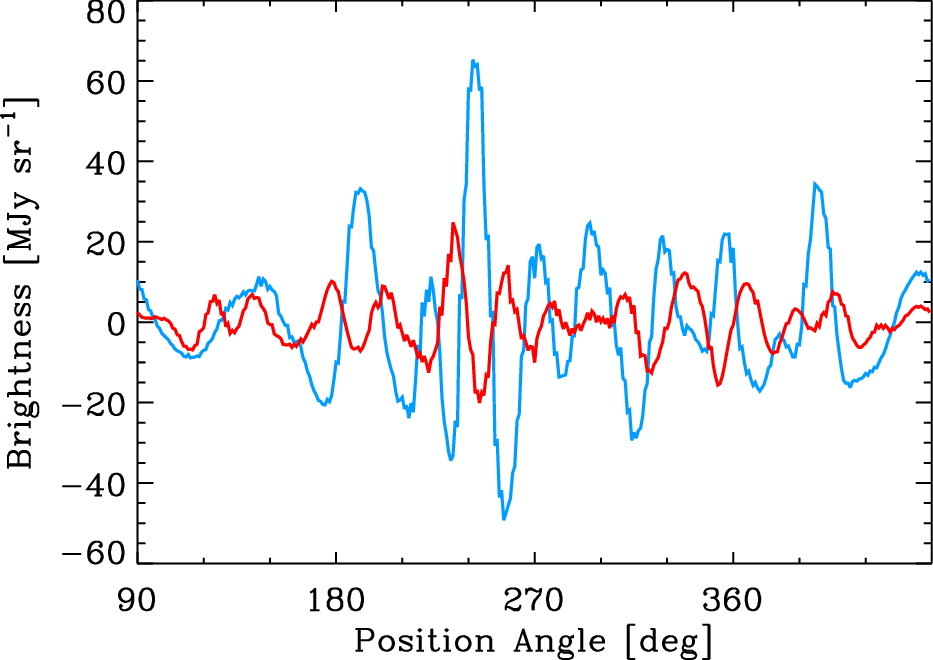}
   \caption{Correlation of 
   the brightness in the inner ER (at F405N) and the outer ER (at F444W).
   Top: Brightness 
   as a function of deprojected position angle 
   along the Figure \ref{fig:er_rings} ellipses tracing the 
   bright hotspots (blue) and the outer emission (red, multiplied by 5). 
   Position angle is measured eastward from north 
   along the ring as if it were viewed face on.
   The dotted lines show the large
   scale variation as traced by the lowest frequency components of the Fourier transform.
   Bottom: The small scale structure of the bright hotspots (blue) and outer emission
   (red, multiplied by 5) after subtraction of the large scale background indicated in the top panel. 
   Spots in the outer portion of the ER are anti-correlated in position angle with the bright 
   hotspots with a correlation coefficient of $-0.38$.}
   \label{fig:er}
\end{figure}

One of the surprises in the NIRCam images is the diffuse outer ER that
lies just outside the bright hotspots of the ER at all wavelengths. 
The structure had not been seen directly in ground-based near-IR imagining
prior to 2017, but comparison with HST data suggests that diffuse IR emission 
began appearing in about 2013 \citep{Kangas:2023}.
There has been evidence of this structure in HST observations \citep{Fransson:2015, Larsson:2019},
but it is far more prominent in the more 
recent and longer wavelength NIRCam data, and is clearly traced by 
\ion{He}{1} emission in the reverse shock \citep{Larsson:2023}. 
An interesting detail of this is shown in
Figures \ref{fig:er_rings} and \ref{fig:er}. The low contrast spots in the diffuse outer ER
are usually found at position angles 
between the bright hotspots of the inner portion of the ER.
This may indicate that the ER hotspots are imprinting small scale structure on 
the reverse shock as it is swept out past the ER. 
Alternately it could be an indication of material (dust) from 
the lower density portions of the ER (or ablated from the hotspots)
that has been entrained in the blast wave that propagated between and around 
the denser hotspots of the ER. This is supported by the fact that the mid-IR 
(silicate) emission is now spatially resolved by both ground-based and
the JWST MIRI
imaging and is 
found to lie in this diffuse region exterior to the bright ER hotspots
\citep{Matsuura:2022,Jones:2023,Bouchet:2023}.
It will be interesting to check the proper 
motion of these features relative to the bright hotspots with future observations.

An outer emission component on the west side of the ER has 
previously been observed with HST in the F502N [\ion{O}{3}]-dominated filter,
suggested to originate from gas swept up by the blast wave propagating in the 
low-density medium between the hotspots \citep{Larsson:2019}. In addition, the HST 
images reveal faint outer spots (mainly in the southeast) and diffuse emission in 
H$\alpha$, which likely originate from high-latitude material.
The high latitude emission from the reverse shock was noted by HST 
in Ly$\alpha$ and H$\alpha$ \citep{Michael:2003,France:2015},
as well as from radio emission \citep{Ng:2008,Fransson:2013,Larsson:2019}.


\section{Spectral Decomposition}

The distinct structures and their colors that are evident in the
NIRCam images \citep[see][and Fig.\ \ref{fig:images}]{Matsuura:2023} suggest that it may be useful to decompose
the spatial structure of the images into separate components, each 
characterized by a different spectrum. Such a decomposition 
can reveal the spatial locations of various emission mechanisms 
and physical conditions, even in places where multiple components 
overlap. This form of spectral decomposition has previously 
been applied to the Cas A SNR \citep{Arendt:2014}.

To perform the spectral decomposition,
at each point [coordinates ($\alpha,\delta$)] in the images the observed SED, $I_{\nu}(\alpha,\delta)$, 
is modeled as the sum of several template SEDs, $F_{\nu,i}$,
representing different emission sources or mechanisms, 
\begin{equation}
I_{\nu}(\alpha,\delta)  = \sum_i A_i(\alpha,\delta) F_{\nu,i}
\end{equation}
where $A_i(\alpha,\delta)$ are the coefficients to be determined at each location.  
With 8 bands, we can solve for no more than 8 free parameters
at each location,
i.e. no more than 8 spectral components.
In the analysis below we
investigate two alternate choices of 5 SED templates, $F_{\nu,i}$ for $i=1,5$.
The parameters $A_i(\alpha,\delta)$ are determined though $\chi^2$ minimization.
However, the $\chi^2$ values do not capture the degree to which the $A_i(\alpha,\delta)$ 
parameters are spatially distinct from one another, and non-negative, 
which are both important considerations here.

We had also applied principal component analysis (PCA) to the NIRCam data.
This does not yield components that are easier to interpret than the original
images, but it does indicate that at least 5 spectral components 
are warranted when modeling the emission in the data set.

\subsection{Decomposition with Sampled SEDs}\label{sec:decomp}

\begin{figure}[t] 
   \centering
   \includegraphics[width=2.5in]{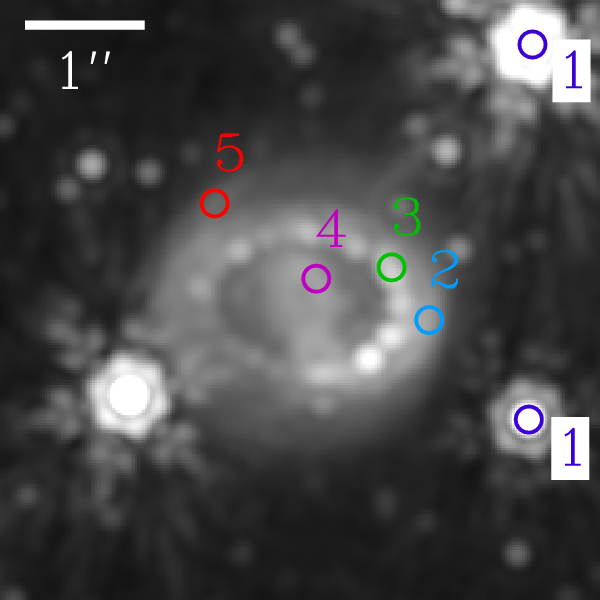} 
   \caption{Locations chosen for empirical SEDs used for spectral decomposition, indicated on the
   F150W image after convolution to F444W resolution. See section \ref{sec:decomp}.
   The crop and display range are the same as Figure \ref{fig:images}.}
   \label{fig:samples}
\end{figure}

For this decomposition we chose five spectral templates that are 
selected from the observed SEDs at representative locations
around SN 1987A, as shown in Figure \ref{fig:samples}. \\
$\bullet$ $F_{\nu,1}$ is the mean SED of Stars 2 and 4, chosen to represent stellar emission 
around the field.\\
$\bullet$ $F_{\nu,2}$ is the SED of the brightest portion of the smooth outer part of the ER, 
chosen to represent this newer development of the ER structure.\\
$\bullet$ $F_{\nu,3}$ is the SED of a bright ER hotspot, representing the older and
more prominent structure of the ER.\\
$\bullet$ $F_{\nu,4}$ is an SED in the northern lobe of the inner ejecta, representative of 
this component.\\
$\bullet$ $F_{\nu,5}$ is the SED at a diffuse arc of emission to the northeast of the ER, which 
represents the developing reverse shock in regions beyond the ER.

The SEDs of the stars and the outer ER
seem to be dominated by continuum emission. The ER hotspot and the ejecta
SEDs have strong line emission components, as evidenced by their
prominence in the F405N and F164N bands. 
These characteristics are
confirmed (at lower spatial resolution) by the NIRSpec observations
of \cite{Larsson:2023}.
The $\sim3400$ km~s$^{-1}$ FWHM of the NIRCam narrow band filters should 
be sufficient to capture the bulk of the line emission from the ER and the 
inner ejecta. Line emission in the reverse shock can appear at velocities
up to $\sim10,000$ km~s$^{-1}$ \citep[e.g.][]{Sonneborn:1998,Michael:1998a,smith:2005,France:2010,Fransson:2013,Larsson:2019},
but comparison of the NIRCam narrow band images with the
corresponding wide band images does not indicate the presence of
significant line emission in the wide bands that is missed by the narrow 
bands due to velocity shifts. In these NIRCam data, the reverse shock 
appears to be dominated by continuum emission.

\begin{figure}[htbp] 
   \centering
   \includegraphics[width=3.25in]{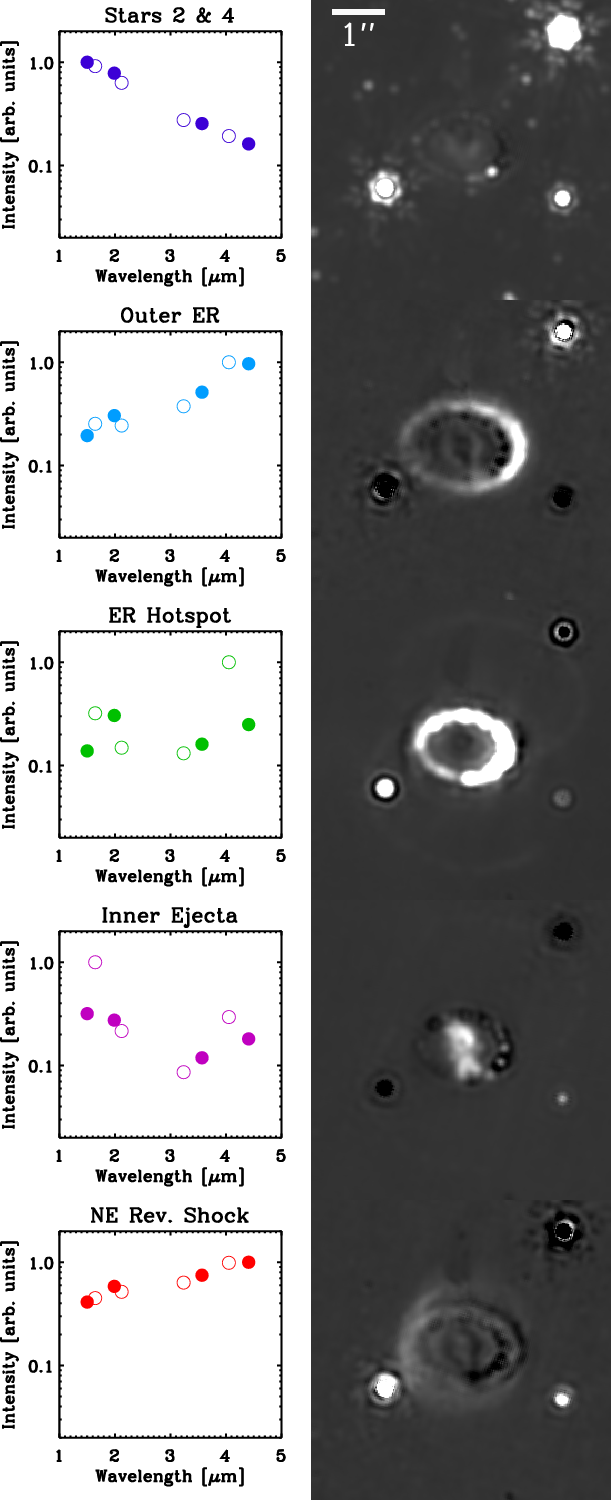} 
   \caption{Spectral decomposition using observed SEDs. The SEDs, $F_{\nu,i}$, (from 
   locations shown in Fig.\ \ref{fig:samples}) are shown in the left column. Wide
   filter bands are indicated with filled circles, narrow ones with open circles.
   The figures in the right column show the derived spatial distribution of the 
   emission, $A_i(\alpha,\delta)$, from each of the SEDs. All are displayed on the same linear scale
   [-5,20] to illustrate the relative importance of the different components.}
   \label{fig:compare321}
\end{figure}

Figure \ref{fig:compare321} shows the five template SEDs and the 
derived spatial distributions of each. (Imperfections in convolving to 
a common resolution leaves ring-like artifacts around bright stars and hotspots.
These should be disregarded.) The emission of stars is largely captured by 
the first template (SED of stars 2 and 4). Note that this template clearly picks out
the faint star that is superimposed on the southwest part of the ER.

The outer ER SED component shows a 
strong gradient (about a factor of 5) in brightness across the ER. This 
component also seems to trace the outer portions of the inner ejecta, which 
suggest that there are spatial variations in the SED of the inner ejecta.
This may arise from asymmetry in the radiation field of the ER and reverse
shock, which heats the inner ejecta, with possible contribution from variations of the 
composition or density of the inner ejecta.

Emission from the ER hotspot SED captures the other hotspots around the inner ER,
which are notably weak in the southeast portion of the ER. This component 
also accounts for much fainter emission from the outer rings of the pre-SN CSM.

The inner ejecta SED clearly defines the spatial distribution of this component,
apparently even in the south where it is beginning to overlap with the ER. The
$A_i(\alpha,\delta)$ for the inner ejecta has moderate negative values 
outlining bright hotspots in the ER. This may
indicate that the ER hotspot SED is not exactly appropriate 
due to changes in the emission between the cores and periphery of the ER hotspots,
or it maybe a result of imperfect PSF matching, as with the 
artifacts around stars.

The appearance of the NE reverse shock component is somewhat similar to 
that of the \ion{He}{1} 1.083 $\mu$m emission that is used to reveal 
the 3-d structure of the reverse shock \citep[Fig. 7 of][]{Larsson:2023}.
[This structure is also traced in H$\alpha$ \citep{Larsson:2019}.]
However in \ion{He}{1}, the emission seems brightest in the NE and SW.
Here, there seems to be east-west gradient that goes in the opposite 
direction as the brightness of the outer ER and the dust temperature 
gradient traced at mid-IR wavelengths 
in the outer ER \citep{Matsuura:2022,Jones:2023,Bouchet:2023}.
On the west side, negative spots are found in the NE reverse shock component 
at bright ER hotspots and bright regions of the outer ER, 
although the structure seems 
heavily influenced by small resolution mismatches.
However, the resulting $A_i(\alpha,\delta)$ image does not match 
well with 315 GHz synchrotron emission mapped by ALMA \citep{Cigan:2019}, 
despite the fact that the NE reverse shock does have a spectral
index very similar to the radio synchrotron emission. The 
map of the 315 GHz emission is best correlated with the outer ER,
which has a much steeper spectrum in the F356W and F444W bands (see below).
  
\subsection{Decomposition with Physical Spectra}
An alternative spectral decomposition is based on SEDs that are 
expected for different emission mechanisms. We again choose
five SED templates.\\
$\bullet$ $F_{\nu,1}$ is derived from a 21000K blackbody spectrum, and is intended to 
represent stars in the field, especially Stars 2, 3, and 4.\\
$\bullet$ $F_{\nu,2}$ is a power law spectrum with a steep spectral index of $\alpha = -3.0$
which is motivated by the general appearance of the spectrum at $\sim 3-10$
$\mu$m \citep{Dwek:2010,Jones:2023} and the specific
ratio of F444W and F356W emission on the west side of the outer ER.
The east side has a flatter spectral index \citep{Matsuura:2023}.
This component represents a combination of continuum emission 
mechanisms, including thermal emission from dust and synchrotron emission
\citep{Jones:2023}. 
The F444W and F356W emissions could alternately be
fit with a blackbody spectrum with $T\approx580$ K. 
The blackbody and power law differ by $<10\%$ at 3-5 $\mu$m,
but the blackbody drops much more sharply at shorter wavelengths.
Neither a power law nor a blackbody would account for the bound-free 
continuum that becomes dominant at $\lambda \lesssim 3 $ $\mu$m \citep{Larsson:2023,Jones:2023}.\\
$\bullet$ $F_{\nu,3}$ is based on integrating a SINgle Faint Object Near-IR Investigation 
\citep[SINFONI;][]{Thatte:1998} line emission spectrum of the ER \citep[data from][]{Larsson:2016} across the
short wavelength NIRCam filter bands, and adding Br~$\alpha$ emission to the 
F405N and F444W bands, assuming a line flux $\sim0.25$ times that measured
in the F200W band (Pa~$\alpha$ + Br~$\gamma$).\\ 
$\bullet$ $F_{\nu,4}$ is a similar integration over a SINFONI line emission spectrum of the ejecta \citep{Larsson:2016}.
In this case no additional lines are included.\\ 
$\bullet$ $F_{\nu,5}$ is intended to represent H$_2$ emission as 
seen in \citep{Fransson:2016,Larsson:2019a,Larsson:2023}.
The model intensities are calculated by integrating the 
\cite{Draine:1996} {\tt Qm3o} photodissociation region (PDR) model, as matched  
to the NIRSpec data \citep{Larsson:2023}, over the NIRCam 
system responses.
This component may be particularly relevant
because the F212N and F323N bands target H$_2$ lines.
In the NIRSpec ejecta spectrum, the 3.23 $\mu$m line is 
much weaker than the 2.12 $\mu$m line \citep{Larsson:2023}.

\begin{figure}[htbp] 
   \centering
   \includegraphics[width=3.25in]{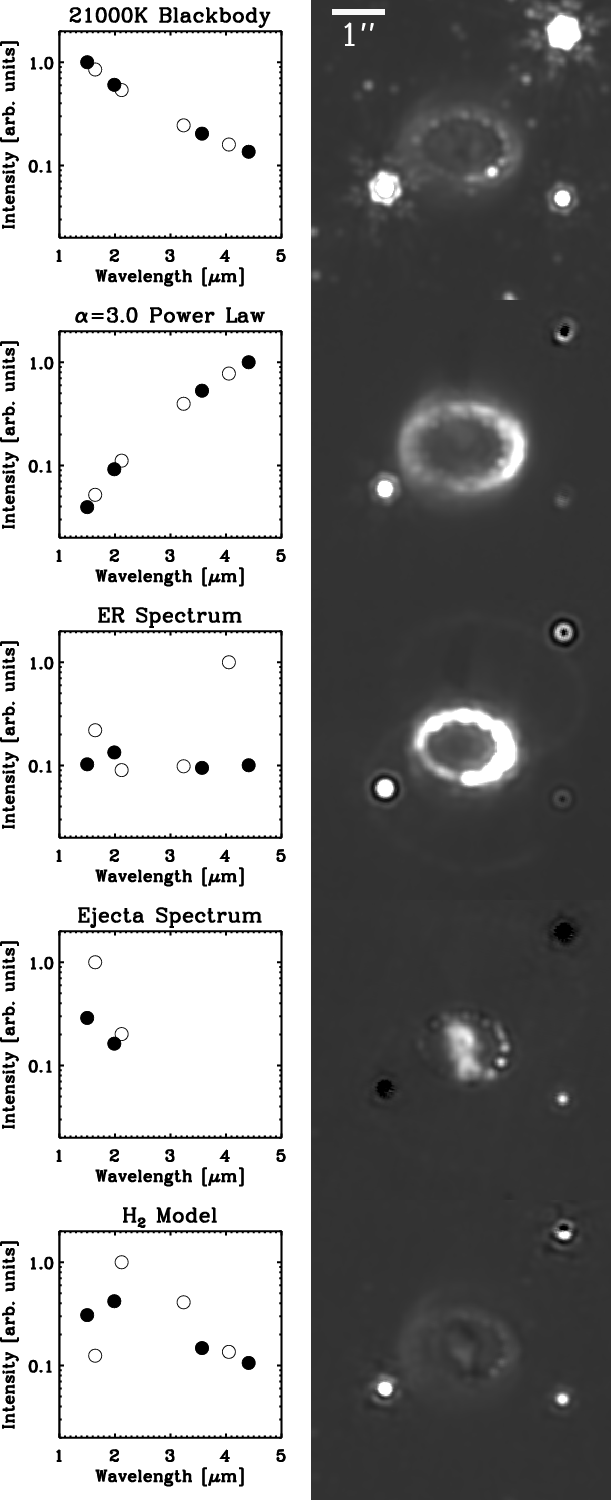} 
   \caption{Spectral decomposition using physically motivated SEDs. The SEDs, $F_{\nu,i}$, 
   are shown in the left column. Wide
   filter bands are indicated with filled circles, narrow ones with open circles.
   The ejecta spectrum SEDs does not have significant emission in all 8 bands. 
   The figures in the right column show the derived spatial distribution of the 
   emission, $A_i(\alpha,\delta)$, from each of the SEDs. All are displayed on the same linear scale
   [-5,20] to illustrate the relative importance of the different components.}
   \label{fig:compare322}
\end{figure}

Figure \ref{fig:compare322} shows these five template SEDs and the 
derived spatial distributions of each. The residuals of the fits 
are somewhat ($\sim3\times$) worse than those of the sampled SEDs from Section 5.1,
but the main results are similar with the following exceptions.
The power law spectrum accounts for both the outer ER and the 
reverse shock regions beyond, which had similar sampled SEDs. 
However since the power law spectrum is relatively weak 
at short wavelengths, the stellar SED is invoked to provide the short wavelength
continuum (mostly bound-free emission) in these regions (especially in the slightly bluer regions
of the reverse shock, e.g. outside the ER to the northeast).
If a 580~K blackbody is substituted for the power law spectrum, then 
this component is similar, except for no longer tracing the hotspots in the ER
which are instead modeled with increased 
contributions from the bluer 20000~K and ejecta components.
The H$_2$ component does isolate the emission of H$_2$ in the ejecta.
This also picks out faint ``crescents'' between the inner 
ejecta and the ER, which are the subject of \cite{Matsuura:2023a}.
However it is relatively weak and also acts as a positive or 
negative correction term to the other templates. 

We also examined this decomposition using the NIRSpec ER and ejecta
spectra from \cite{Larsson:2023}. While these data have complete coverage 
of the NIRCam bands, and better sensitivity than the SINFONI data, 
they are less suitable for this analysis because the published spectra 
are integrated over large regions where multiple emission components overlap.
The ER spectrum, for example, includes the bright hotspots as well as the 
diffuse outer ER and regions of the RS. Extracting spectra at more localized
regions that are dominated by specific emission components leads to 
better SEDs for use in the spectral decomposition of the NIRCam data.

\section{Summary} \label{sec:results}

We have used the high sensitivity and spatial resolution JWST NIRCam $1.5-4.5$ $\mu$m imaging to show consistency with the marginally-resolved Spitzer IRAC images. We confirmed that the Spitzer 3.6 and 4.5 $\mu$m emission arises from the ER, and showed that integrated NIRCam flux densities from the ER in the 3.6 and 4.5 $\mu$m bands fall on the predicted extrapolations of the Spitzer 3.6 and 4.5 $\mu$m light curves. The extrapolations employ a model in which the light curves are fitted by a convolution of a Gaussian with a decreasing exponential function. The two components may represent the convolution of a
(Gaussian) physical distribution of swept up dust with a sharp onset and exponential decay in the emissivity of the dust, or conversely, 
an exponentially declining spatial distribution of dust with a
rising and falling (near Gaussian in shape) emissivity in response to the passing shock.
Such mathematical presentation of the evolution of the light curves may not be unique, however the predictive power of the present one remains compelling.

The F356W flux densities of the companion stars are similar to those modeled 
from the lower resolution Spitzer images. However, we have not applied the newly-measured 
F356W and F444W flux densities retroactively to the Spitzer data because the differences 
from the previously assumed values are relatively small, and because the potential 
variability of Star 3 adds a similar level of uncertainty.

The NIRCam images allow a detailed analysis of the spatial correlation between the ER hotspots and the diffuse extended emission outside the ER. We find an anticorrelation between the azimuthal distribution of the bright inner hotspots and clumps in the outer diffuse emission. The hotspots seem to leave an imprint on the diffuse emission. One possibility is that the diffuse emission represents the forward shock draping around and extending beyond the hotspots \citep[e.g.][]{Silvia:2010,silvia:2012,Kirchschlager:2019,Kirchschlager:2023}. 
The diffuse outer ER seems to coincide with the region where the dominant $10-30$ $\mu$m emission from silicate dust arises \citep{Jones:2023,Bouchet:2023}.  

We used the high spatial resolution images of the SN and the companion stars to decompose and represent the emission 
by two sets of 5 standard SEDs. One set of these SEDs is chosen from 5 distinct representative locations 
in the images. The other set of SEDs is derived from theoretical (blackbody, synchrotron, H$_2$) and empirical 
(ER, ejecta) spectra. These decompositions show that the bulk of the emission can be attributed to only 4 spectrally and spatially 
distinct components of the SN (plus a fifth stellar component): the bright inner ER hotspots, the redder and more diffuse outer ER, 
the inner ejecta, and the reverse shock. The slightly incomplete spatial separation of the components indicates that 
inner ejecta shows the most evidence of distinct spatial variations in its SED.

Our analysis of the high-resolution NIRCam SN~1987A images provides a powerful method for
dissecting the various components of this extremely young SNR. It separates CSM
features excited by the passage of the forward shock, the outer ejecta of the SN
passing through the reverse shock, and the metal-rich inner ejecta which is only
starting to reach the reverse shock. Future combination of the NIRCam imaging
with data from Hubble (with comparable angular resolution) will allow the
distinction of spectral variations within these components and/or additional
components. The spectral decomposition applied to the JWST NIRSpec and MIRI MRS
data cubes \citep{Larsson:2023,Jones:2023} can provide much better definition of
the spectral properties of each component, despite the poorer spatial resolution
of those data.

\begin{acknowledgments}
This work is based on observations made with the 
NASA/ESA/CSA James Webb Space Telescope. The data were
obtained from the Mikulski Archive for Space Telescopes at the Space 
Telescope Science Institute, which is operated 
by the Association of Universities for Research in Astronomy, Inc., 
under NASA contract NAS 5-03127 for JWST. These
observations are associated with program \#1726.
Support for program \#1726 was provided by NASA through a grant from the Space 
Telescope Science Institute, which is operated by the Association of Universities 
for Research in Astronomy, Inc., under NASA contract NAS 5-03127.
Work by R.G.A. was supported by NASA under award number 80GSFC21M0002. 
I.D.L. and F.K. acknowledge funding from the European Research Council (ERC) under the European Union's Horizon 2020 research and innovation programme (\#851622 DustOrigin).
C.G. is supported by a VILLUM FONDEN Villum Young Investigator Grant (25501).
R.D.G. was supported, in part, by the United States Air Force.
P.L. thanks the Swedish Research Council for support.
M.M. and R.W. acknowledge support from STFC STFC Consolidated grant (2422911).
J.C.W. is supported by NSF Grant AST 1813825. 
We thank the anonymous referee for helpful advice on improvements to the text and figures.
\end{acknowledgments}

\facilities{JWST(NIRCam)}

\software{IDLASTRO \citep{Landsman:1995}, JHAT \citep{Rest:2023}}

\bibliographystyle{aasjournal}
\bibliography{decomp_sn1987A}

\end{document}